\documentclass[10pt,journal]{IEEEtran}
\usepackage{amsmath}
\usepackage{amsfonts}
\usepackage{amssymb}
\usepackage{array}
\usepackage{multirow}
\usepackage{multicol}
\usepackage{graphicx}
\usepackage{subfig}
\usepackage{threeparttable}
\usepackage{color}
\usepackage{soul}

\usepackage{algorithm}
\usepackage[noend]{algpseudocode}
\usepackage{threeparttable}

\begin{document}

\title{A Bayesian incorporated linear non-Gaussian acyclic model for multiple directed graph estimation to study brain emotion circuit development in adolescence}

\author{Aiying~Zhang,
        Gemeng~Zhang,
        Biao~Cai,
        Tony~W.~Wilson,~\IEEEmembership{Senior Member,~IEEE},
        ~Julia~M.~Stephen,~\IEEEmembership{Senior Member,~IEEE},
        ~Vince~D.~Calhoun,~\IEEEmembership{Fellow,~IEEE},
        and Yu-Ping~Wang,~\IEEEmembership{Senior Member,~IEEE}
\thanks{A. Zhang, G. Zhang, B. Cai and Y.-P. Wang are with the Department of Biomedical Engineering, Tulane University, New Orleans, LA 70118, USA e-mail: wyp@tulane.edu.}
\thanks{T. W. Wilson is with the Department of Neurological Sciences, University
of Nebraska Medical Center, Omaha, NE 68198 USA.}
\thanks{J. M. Stephen and V. D. Calhoun are with Mind Research Network,
Albuquerque, NM 87106 USA.}
\thanks{V. D. Calhoun is with Tri-institutional Center for Translational Research in Neuroimaging and Data Science (TReNDS), Georgia State University, Georgia Institute of Technology, Emory University, Atlanta, GA, 30303 USA}
}

\maketitle

\begin{abstract}

Emotion perception is essential to affective and cognitive development which involves distributed brain circuits. The ability of emotion identification begins in infancy and continues to develop throughout childhood and adolescence. Understanding the development of brain's emotion circuitry may help us explain the emotional changes observed during adolescence. Our previous study delineated the trajectory of brain functional connectivity (FC) from late childhood to early adulthood during emotion identification tasks. In this work, we endeavour to deepen our understanding from association to causation. We proposed a Bayesian incorporated linear non-Gaussian acyclic model (BiLiNGAM), which incorporated our previous association model into the prior estimation pipeline. In particular, it can jointly estimate multiple directed acyclic graphs (DAGs) for multiple age groups at different developmental stages. Simulation results indicated more stable and accurate performance over various settings, especially when the sample size was small (high-dimensional cases). We then applied to the analysis of real data from the Philadelphia Neurodevelopmental Cohort (PNC). This included 855 individuals aged 8–22 years who were divided into five different adolescent stages. Our network analysis revealed the development of emotion-related intra- and inter- modular connectivity and pinpointed several emotion-related hubs. We further categorized the hubs into two types: in-hubs and out-hubs, as the center of receiving and distributing information. Several unique developmental hub structures and group-specific patterns were also discovered. Our findings help provide a causal understanding of emotion development in the human brain.
\end{abstract}


\section{Introduction}

Emotion perception is essential to affective and cognitive development and is thought to involve distributed brain circuits. Identification of distinct facial expressions, which is fundamental to recognizing the emotional state of others, begins in infancy and continues to develop throughout childhood and adolescence \cite{zhang2019development}. Understanding the development of brain's emotion circuits may help us explain the emotional maturation seen in adolescence. In our previous work \cite{zhangTMI2019}, we studied the fMRI images collected on the Philadelphia Neurodevelopmental Cohort (PNC) and delineated the trajectory of the brain functional connectivity (FC) from late childhood (preadolescence) to early adulthood (post-adolescence) during emotion identification tasks. The FC metrics that we used were defined by statistical associations (partial correlations, in specific) between measured brain regions. However, it has been pointed out that the statistical association may be problematic in that it only reveals the spatial connections but not causal information \cite{reid2019}. Approaches that characterize statistical associations are likely a good starting point, but the causality of brain connectivity should be more promising.

Directed acyclic graph (DAG) models, also known as Bayesian networks, are designed to model causal relationships in complex systems. Current methods for DAG identification can be divided into four categories: constraint-based methods, score-based methods, non-Gaussian based methods, and hybrids of these categories. The constraint-based methods, such as the PC algorithm (named after its authors, Peter and Clark) \cite{pc2010}, and the score-based methods, such as the greedy equivalent search (GES) have been found sub-optimal for predicting the direction of causal relationships but are accurate in identifying the causal skeleton (graph structure without directions) \cite{smith2011network}. The non-Gaussian based methods, the linear non-Gaussian acyclic models (LiNGAMs), have better performance since the non-gaussian data may contain more information to infer the directionalities \cite{Ramsey2014}. However, it requires a large number of data points in the relevant dimension to converge to the true graph \cite{smith2011network}. A recent review discussed that by leveraging insights from existing association studies, we can reduce the set of likely causal models, facilitating causal inferences despite major limitations \cite{reid2019}. Methods like high-dimensional PC \cite{kalisch2007estimating}, fast GES \cite{ramsey2017million} and $\psi$-LiNGAM \cite{zhangOHBM2020} that incorporated association networks have all shown great improvements.

Currently, existing methods have focused on estimating a single directed graphical model.
However, in many biomedical applications, we have access to data from related classes, like the multiple adolescent periods in our adolescence study. This raises an important statistical question, namely how to jointly estimate related graphical models in order to make full use of the available data \cite{wang2018}. We thus developed a joint Bayesian-incorporating $\psi$-learning method to estimate multiple undirected graphical models as priors based on our previous study \cite{Zhang2018} and proposed a multiple DAG estimation for non-Gaussian data. We named it the Bayesian incorporated linear non-Gaussian acyclic model (BiLiNGAM) in the sense that we were inspired by the idea of Bayesian prior and implemented a joint Bayesian-incorporating estimation to acquire the priors. A series of simulation studies have been conducted to further illustrate the advantages of BiLiNGAM in terms of convergence speed and accuracy, especially in high dimensional datasets. Network analysis of brain connectivity development using the fMRI image from PNC revealed the development of emotion-related intra- and inter- modular connectivity and pinpointed a few emotion-related hubs. We further categorized them into two types: in-hubs and out-hubs, as the center of receiving and distributing information. Some unique developmental hub structures and group-specific patterns have also been discovered.

The remainder of this paper is organized as follows. In Section \uppercase\expandafter{\romannumeral2}, we introduce the BiLiNGAM method step by step. A series of simulation studies are displayed in Section \uppercase\expandafter{\romannumeral3}. A detailed analysis of causal emotion circuit development of human brain in adolescence is illustrated in Section \uppercase\expandafter{\romannumeral4}, followed by some concluding remarks in the last section.

\section{Methods}

In this section, we introduce the Bayesian incorporated linear non-Gaussian acyclic model (BiLiNGAM) step by step. The proposed model is developed from the LiNGAM method and incorporate prior information for multiple groups that are distinct but related. We first introduce the general LiNGAM methods in Section \uppercase\expandafter{\romannumeral2}.A. Section \uppercase\expandafter{\romannumeral2}.B briefly describes the approach for prior knowledge estimation for multiple groups. Finally, we summarize the BiLiNGAM algorithm in Section \uppercase\expandafter{\romannumeral2}.C.

\subsection{LiNGAM}

The linear non-Gaussian acyclic model (LiNGAM) was first proposed to study Bayesian networks (BN) using a structural equation model (SEM) for non-Gaussian variants \cite{Shimizu2006}. LiNGAM assumes that the casual relationships of the variables can be represented graphically by a directed acyclic graph (DAG) $G = (V, \bold{E})$, where the node set $V = \{1,2,...,p\}$ represents the corresponding variables and $\bold{E} \in R_{p \times p}$ denotes the adjacency matrix of the directed edges. Let $\bold{B} =\{b_{ij}\} \in R^{p \times p}$ be the weighted adjacency matrix specifying the edge weights of the underlying DAG $G$. The observed random vector $\bold{X} = (\bold{X}_1, \bold{X}_2, ..., \bold{X}_p) \in R^p$ is assumed to be generated from the following linear SEM,
\[
\bold{X} = \bold{B}^T \bold{X} + \boldsymbol{\epsilon},
\]
where $\boldsymbol{\epsilon} = (\boldsymbol{\epsilon}_1, \boldsymbol{\epsilon}_2, ..., \boldsymbol{\epsilon}_p)$ is a continuous random vector; the $\boldsymbol{\epsilon}_i$'s, $\forall i =1,2,..., p$ have non-Gaussian distributions with non-zero variances, and are independent of each other.

A property of acyclicity is that there exists at least one permutation $\pi$ of $p$ variables such that $b_{ij} = 0$, $\forall \pi(i) < \pi(j)$. In other words, the weight matrix $\bold{B}$ can be reordered to a strictly lower triangular matrix according to the permutation $\pi$. The goal of LiNGAM is to find the correct permutation and estimate the weight matrix $\bold{B}$. Since the components of $\boldsymbol{\epsilon}$ are independent and non-Gaussian, Shimizu et al. \cite{Shimizu2006} first proposed the independent component analysis (ICA) based algorithm known as ICA-LiNGAM. Later, another method, the direct LiNGAM \cite{Shimizu2011}, has been developed which estimates the causal order of variables by successively subtracting the effect of each independent component from given data. Compared to the ICA-based algorithm, the direct LiNGAM needs no initial guess or algorithmic parameters and has guaranteed convergence.

LiNGAM is designed for non-gaussian data, which contains more orientation information. Therefore it is possible to identify more of the causal graph structure than the traditional Gaussian setting \cite{Ramsey2014}. However, the identification of DAG models in general is nondeterministic polynomial time hard (NP-hard) \cite{Heckerman1995}. LiNGAM, especially, requires a larger number of data points in the relevant dimension to converge to the true graph \cite{smith2011network}. In biomedical applications, the datasets collected often have limited sample size or the problems we are facing are high dimensional cases (i.e., the number of variables/nodes greatly exceeds the number of samples/observations). Under these circumstances, the traditional LiNGAM methods could give questionable results or cannot even be applied to high dimensional cases.

\subsection{Bayesian Incorporated Prior Estimation}

Although investigating causal interactions should be central to studies of complex systems like the brain function, associations are a good starting point for estimating network interactions and various methods have been proposed. Current methods can be categorized into three groups based on the statistical associations to be calculated, which, namely, are the Pearson correlations, partial correlations and distance correlations. Pearson correlation describes the linear correlation of a pair of variables in a system, partial correlation measures the association between two variables removing the effect of other variables, and distance correlation can measure both linear and nonlinear association between two random variables. Based on \cite{reid2019}, generalizing insights from existing association studies can facilitate causal inferences and overcome major limitations such as small sample size and computational inefficiency. As in statistical modelling, we can incorporate the established networks as prior information with existing causal models. Within a linear model, a partial correlation-based approach is more appropriate than the one using Pearson correlation. This is because in a complex system like the brain, the Pearson correlation is much weaker marginally \cite{Liang2015}, i.e., all nodes (variables) are directly or indirectly correlated and it is difficult to distinguish significant connections through a dense network constructed by Pearson correlations. Therefore, incorporating the Pearson correlation network as a prior may bring too much confounding information and the improvement of causal inferences is limited. On the contrary, partial correlations have been proposed to explore direct associations between two nodes, controlling for the confounding variables, which can help to infer causal interactions. In our previous causal study \cite{zhangOHBM2020}, we incorporated the partial correlation network using the $\psi$-learning method \cite{Liang2015} as a prior with LiNGAM and showed its superiorities of convergence speed and accuracy.

Our goal in this paper is to jointly estimate multiple related DAGs. Therefore, we developed a joint Bayesian-incorporating $\psi$-learning method to estimate multiple undirected graphical models as priors. The approach consisted of three steps: Step 1, Gaussian transformation; Step 2, distinct and common graph construction; and Step 3, prior matrices acquisition. Plenty of the partial correlation based methods have been working on the Gaussian distributed data due to their mathematical simplicity \cite{Zhang2018}. Liu et al. \cite{Liu2009} have proposed a nonparanormal transformation which relax the Gaussian assumption to continuous. Thus in Step 1, we apply this Gaussian transformation. For Step 2, we take into consideration the distinct and common structure for each group. The distinct graph estimation for each group is implemented through the $\psi$-learning method as in \cite{Zhang2018}. To strengthen the similarities over various groups, we adopt the same Bayesian incorporating joint estimation method as in our previous study of brain connectivity development in adolescence \cite{zhangTMI2019}. The similarities are highlighted through proper Bayesian priors and a meta-analysis procedure. Finally, we use the union of the distinct estimated graph and joint estimated graph as the prior graph for each group.

\subsection{BiLiNGAM}

In the literature, most existing methods have focused on estimating a single directed graphical model. For multiple DAG estimation, Wang et al. \cite{wang2018} have proposed a high-dimensional joint estimation of multiple directed graphical models for Gaussian distributed data. In this paper, we propose a multiple DAG estimation for non-Gaussian data, with the name of Bayesian incorporated linear non-Gaussian acyclic model (BiLiNGAM) in the sense that we were enlightened by the idea of Bayesian prior and implemented a joint Bayesian-incorporating estimation to acquire the priors. The procedure of the BiLiNGAM algorithm is summarized in Algorithm 1.

\algsetblock[Name]{Start}{End}{4}{0.2cm}
\renewcommand{\algorithmicrequire}{\textbf{Input:}}
\renewcommand{\algorithmicensure}{\textbf{Output:}}
\renewcommand{\algorithmicprocedure}{\textbf{Procedure:}}
\begin{algorithm}
\caption{BiLiNGAM algorithm}\label{BiLiNGAM}
\begin{algorithmic}
\Require Collection of observations $\bold{X}^k = (\bold{X}_i^k) \in \mathbb{R}^{n_k \times p}$, where $k = 1,2,\dots, K$, $i=1,2,\dots,p$ and $\bold{X_i^k}$'s are non-Gaussian continuous.
\Ensure Collection of estimated weighted adjacency matrices $\hat{B}^k$
\State 1. Prior estimation: joint Bayesian-incorporating $\psi$-learning.
\Start:
  \State a. For $k=1,2,\dots,K$, use the nonparanormal transformation \cite{Liu2009} to render $\bold{X}^k$ normal (Gaussian).
  \State b. Apply the $\psi$-learning method \cite{Liang2015} to each group $k$, $k = 1,2,\dots, K$ separately for distinct estimation and acquire the adjacency matrix $\bold{E}^{d,k}$.
  \State c. Apply the Bayesian incorporating joint estimation \cite{zhangTMI2019} to strengthen the similarities among the groups and acquire the $\bold{E}^{c,k}$, $\forall k$.
  \State d. Extract the prior matrix $\bold{A}^{prior,k}$ from $\bold{E}^{prior,k} = \bold{E}^{c,k} \cup \bold{E}^{d,k}$, where $a_{ij}^{prior,k} = -1$, if $ e_{ij}^{prior,k} = 1$ and otherwise $a_{ij}^{prior,k} = 0$.
\End

\State 2. Obtain the estimated weighted DAG adjacency matrices $\hat{\bold{B}}^k$: LiNGAM.
\Start:  For each $k$
\State a. Identify the casual order $\pi^k$ using the direct LiNGAM with the prior matrix $\bold{A}^{prior,k}$ \cite{Shimizu2011}.
\State b. Construct a strictly lower triangular matrix $\tilde{\bold{B}}^k$ by following the causal order $\pi^k$, and the corresponding $\tilde{\bold{A}}^{prior,k}$ with the same order.
\State c. Estimate the connection strengths $(\tilde{\bold{B}}_{j}^k)^T = (\tilde{b}_{1j}^k, \tilde{b}_{2j}^k, ..., \tilde{b}_{pj}^k )$ consistent with $\tilde{\bold{A}}^{prior,k}$ by solving sparse regressions of the form
\[
\hat{\tilde{\bold{B}}}_j^k = \arg \min \limits_{\tilde{\bold{B}}_j^k \subset supp(\tilde{a}_j^{prior,k})} ||\bold{X}_j^k - \bold{X}^k \tilde{\bold{B}}_j^k||_2^2
\]
\State d. Obtain $\hat{\bold{B}}^k$ by converting $\tilde{\bold{B}}^k$ to the original order.

\End

\end{algorithmic}
\end{algorithm}

\section{Simulation Studies}

In this section, we analyzed the performance of the joint estimation of $K$ different DAGs where we varied $K \in \{3, 5\}$. For all experiments, we set the number of nodes $p = 200$ and the total number of observations $N = 750$. For each group, we set the number of samples equally, i.e. $n_1 = n_2 = \dots = n_K = n = N/K$. The random DAG $G$ can be simulated through the R package \emph{pcalg} and density of the graph is controlled by the edge probability $d/(p-1)$, where $d$ is an edge degree parameter with values $\{1, 2, 5 \}$. The true DAGs generation procedure is illustrated as follows. We first used the \emph{pcalg} to generate $G^1$. Given $G^1$, we assigned uniformly random weights to the edges to obtain the weighted adjacency matrix $\bold{B}^1 = (b_{ij}^1)$: $b_{ij}^1 \sim \rm Unif (-0.8,-0.3)\cup (0.3,0.8)$, if there is an edge $i \rightarrow j$, otherwise $b_{ij}^1=0$. For $G^k$, $k=2, 3, \dots, K$, we followed the same random edge deleting-adding procedure in a sequential manner. We randomly removed $5\%$ edges in $G^{k-1}$, $k=2,...,5$, by setting the corresponding non-zero elements in $\bold{B}^k$ to be $0$, and then added $5\%$ edges at random by giving them values drawn from the uniform distribution $U[0.3,0.5]$ to obtain $\bold{B}^k$. Given $\bold{B}^k$'s, we generated $\bold{X}^k = (\bold{B}^k)^T \bold{X}^k + \boldsymbol{\epsilon}^k \in R^p$ with $\boldsymbol{\epsilon}^k$ from Chi-squared (Chisq) noise with degree of freedom $1$ and zero mean, i.e. $\epsilon_i^k \sim \chi^2_1 - 1$, $i,=1, 2, \dots, p$, $k = 1,2, \dots, K$.

We considered three methods for comparison, which are the PC algorithm \cite{pc2010}, the ICA-LiNGAM \cite{Shimizu2006} and the $\psi$-LiNGAM we proposed previously \cite{zhangOHBM2020}. The prominent PC ("Peter and Clark") algorithm \cite{pc2010} is a type of constraint-based methods, that first learn an undirected graph from conditional independence relationships, which is called as the skeleton of the directed graph, and then orient the edges. Studies \cite{smith2011network, henry2017causal} have shown that the PC algorithm is good at the skeleton estimation but not the edge orientation. The PC and ICA-LiNGAM were implemented through the R package \emph{pcalg} and the code for $\psi$-LiNGAM is available at https://github.com/Aiying0512/psi-LiNGAM. We set the significance level $\alpha = 0.05$ with FDR correction for the PC, $\psi$-LiNGAM and BiLiNGAM. For each scenario, 10 datasets were simulated independently. We assessed the performances of the four methods through the true positive rate (TPR), false discovery rate (FDR), and structural hamming distance (SHD) \cite{SHD2006}. TPR and FDR are two common measures of binary classification. Let us define an experiment from $P$ positive instances and $N$ negative instances for some conditions. In our case, the positive instance represents a directed edge from one node to the other. The four outcomes are summarized in TABLE \uppercase\expandafter{\romannumeral1}.
The definitions of TPR and FDR are given as follows:
\[
\rm TPR = \frac{TP}{TP+FN}, \quad FDR = \frac{FP}{FP+TP}.
\]

\linespread{1.25}
\begin{table}[h!]
\centering
\caption{Outcomes of a binary decision}
\begin{tabular}{ccc}
\hline
                  & Actual positive ($P$) & Actual negative ($N$) \tabularnewline
\hline
Predicted positive & True positive (TP) & False positive (FP)\tabularnewline
Predicted negative & False negative (FN) & True negative (TN)\tabularnewline
\hline

\end{tabular}
\end{table}

\begin{figure}[h!]
\centering
\includegraphics[width=3.5in]{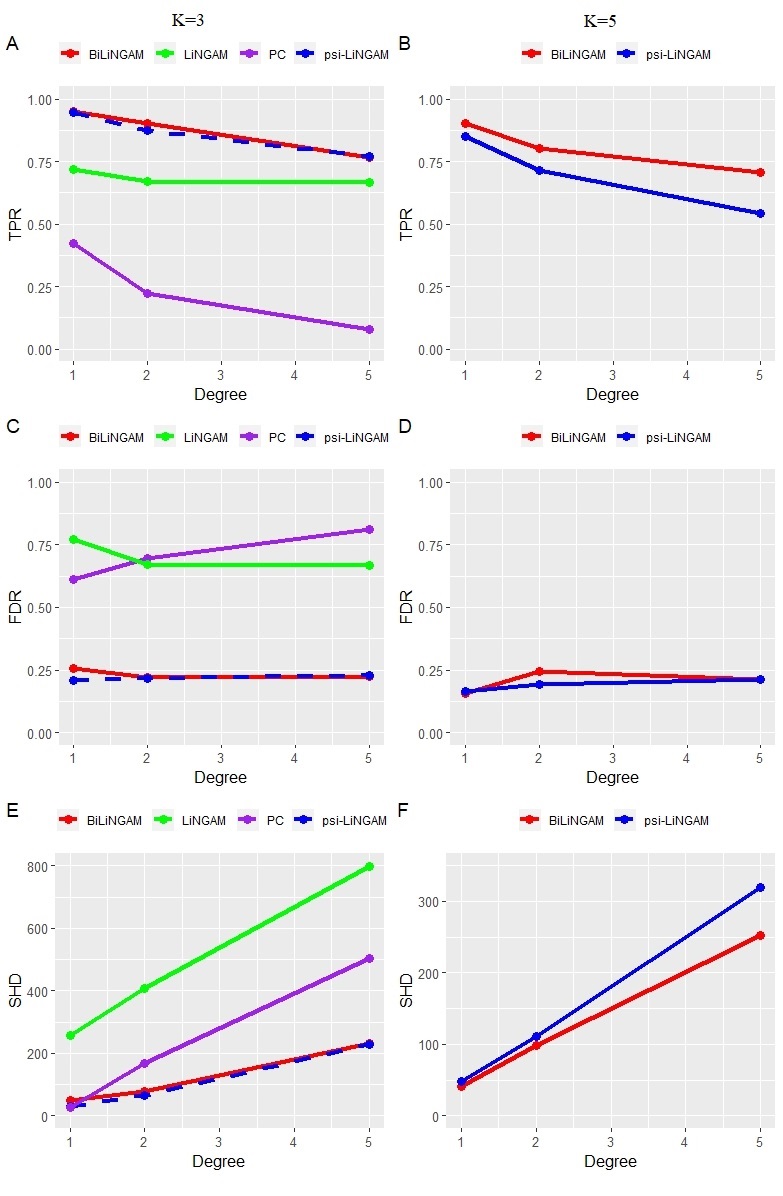}
\caption{Simulation results of the mean TPR, FDR and SHD over various graph densities. The performances of $4$ methods for $K=3$ are on the left column (A, C, E). The comparisons between BiLiNGAM and $\psi$-LiNGAM for $K=5$ are on the right column (B, D, F). }\label{fig:1}
\end{figure}

SHD is a frequently used metric based on the number of operations needed to transform the estimated DAG into the true graph \cite{Kalisch2007}. In simple terms, SHD counts the total number of edge insertions, deletions or flips during the transformation. Fig. \ref{fig:1} gives the results for $K=3$, where each group has $n = 250$ samples and $p = 200$ nodes. We compared the results of all four methods mentioned above with various graph densities controlled by the degree parameter $d$. The results from Fig. \ref{fig:1} (left column) validate the poor edge orientation ability of the PC algorithm. Although the TPR curve of ICA-LiNGAM indicates a decent performance, the high FDR and SHD prove that the ICA-LiNGAM need a larger number of observations to converge to the true graph. As we can see, the two methods incorporating association networks as prior information ($\psi$-LiNGAM and BiLiNGAM) performs similarly well. This shows that when the sample size is sufficient ($n>p$), both methods incorporating association networks to estimate multiple DAGs can improve causal inferences equivalently. Further, we compared the two methods under high dimensional cases (i.e., $n<p$). Fig. \ref{fig:1} (right column) shows the results for $K=5$, where each group has $n=150$ samples. The FDR curves of the two methods maintain at the same low level. The TPR of BiLiNGAM is always higher than the $\psi$-LiNGAM's and the SHD performs the opposite. In addition, as the degree parameter $d$ increases, the differences in TPR and SHD also increase. Therefore, under high dimensional settings, both methods remain at a low FDR level, but BiLiNGAM outperforms $\psi$-LiNGAM in terms of TPR and SHD. Overall, BiLiNGAM has maintained a stable and accurate performance over various settings. Particularly, under high dimensional cases, the performance of BiLiNGAM is superior. When the sample size is adequate (i.e. $n>p$), BiLiNGAM performs at least as good as $\psi$-LiNGAM.

%

\section{A Study of Adolescent Brain Connectivity Development}

\subsection{Materials}

\begin{figure}[h!]
\centering
\includegraphics[width=3.5in]{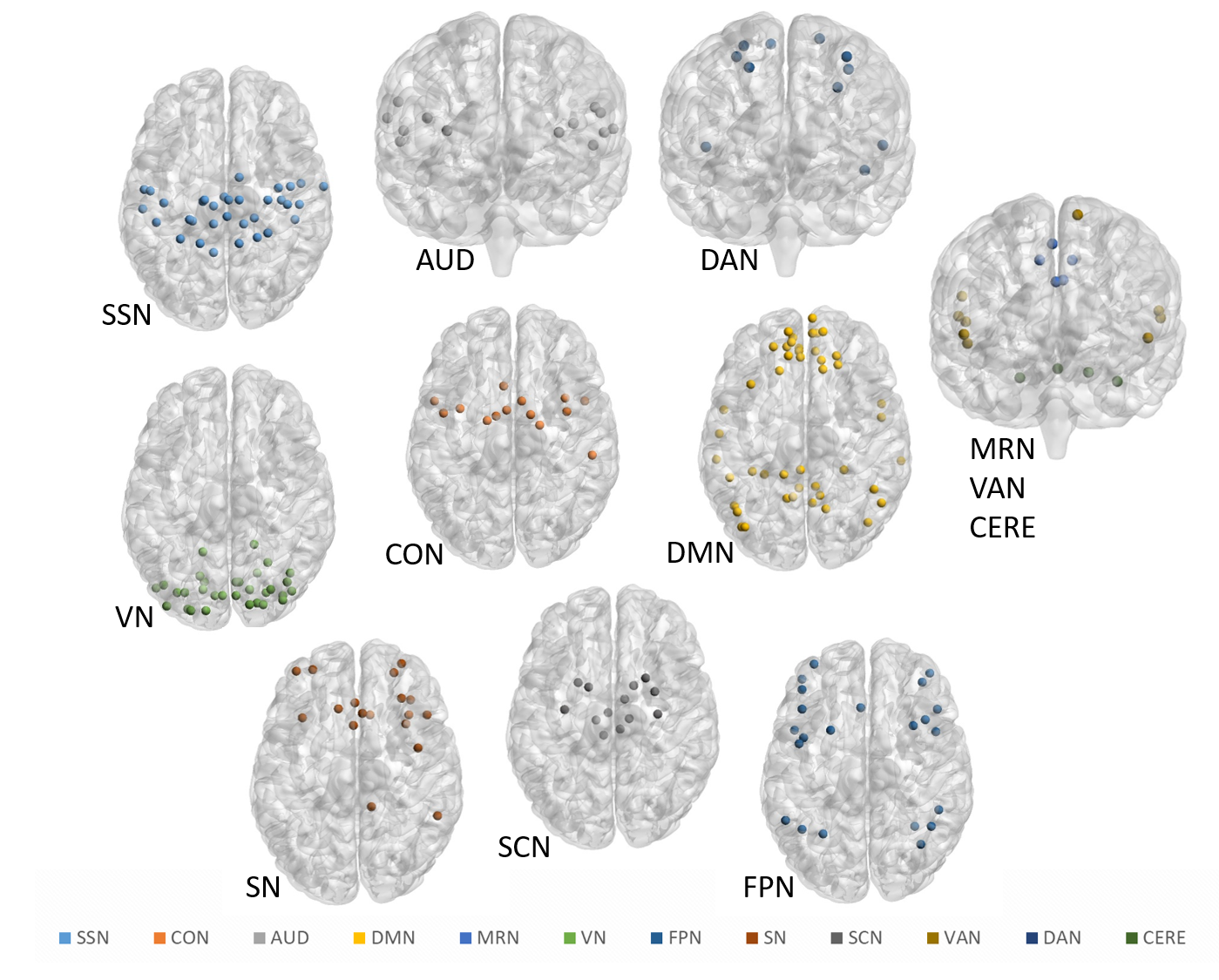}
\caption{12 functional network modules based on the 264 nodes from the template defined by Power et al. \cite{Power2010}.}\label{fig:2}
\end{figure}

The dataset we used is publicly available from the Philadelphia Neurodevelopmental Cohort (PNC). It consists of fMRI images from 855 individuals using an emotion identification task. All MRI scans were acquired on a single 3T Siemens TIM Trio whole-body scanner. During the task, each subject was asked to label emotions displayed which include happy, angry, sad, fearful and neutral faces. The total scan duration was 10.5 min. Blood oxygenation level-dependent (BOLD) fMRI was acquired using a whole-brain, single-shot, multislice, gradient-echo (GE) echoplanar (EPI) sequence of 124 volumes (372s) with the following parameters TR/TE=3000/32 ms, ﬂip $=90^\circ$, FOV$=192 \times 192$ mm, matrix $= 64 \times 64$, slice thickness/gap=3 mm/0 mm. The resulting nominal voxel size was $3.0 \times 3.0 \times 3.0 $mm \cite{Satt2014}. Standard preprocessing steps were applied using SPM12, including motion correction, spatial normalization to standard MNI space, and spatial smoothing with a 3mm full width at half max (FWHM) Gaussian kernel. Then multiple regression considering the influence of motion was performed and the stimulus on-off contrast maps for each subject were obtained. Finally, 264 functionally defined regions of interest (ROIs) were extracted based on the Power parcellation \cite{Power2010}. The age range of the participating subjects was between 8 and 22 years. Due to physical and cognitive changes \cite{zhangTMI2019}, we divided them into five groups, each representing a stage related to adolescence (Table \uppercase\expandafter{\romannumeral2}) as shown in Fig. \ref{fig:2}.

\begin{table}[h]
\caption{Group division information.}
\centering
\begin{tabular}{|c|c|c|c|}
\hline
Category &      Group   & Age    &  Number of \\
index    &      name    & range  &  subjects \\
\hline
1 &     Pre-adolescence  & 8-12  & 194\\
2 &     Early adolescence &  12-14   & 150\\
3 &     Middle adolescence & 14-16 &158\\
4 &     Late adolescence &  16-18 & 166\\
5 &     Post-adolescence &  18-22 & 187\\
\hline
\end{tabular}
\end{table}

Furthermore, we divided the $264$ brain regions into 12 functional network (FN) modules to study regional connectivity, including sensory/somatomotor network (SSN), cingulo-opercular task control network (CON), auditory network (AUD), default mode network (DMN), memory retrieval network (MRN), visual network (VN), fronto-parietal task control network (FPN), salience network (SN), subcortical network (SCN), ventral attention network (VAN), dorsal attention network (DAN) and cerebellum network (CERE).

\subsection{Results}

We first conducted the Darling-Anderson test for non-Gaussianity and then applied BiLiNGAM. The causal brain connectivity of each group is shown in Fig \ref{fig:3} A. We summarized the networks from the aspects of modular development and significant ROIs (hubs).

\subsubsection{Development of emotion-related intra- and inter- module connectivity}

\begin{figure*}[h!]
\centering
\includegraphics[width=5.4in]{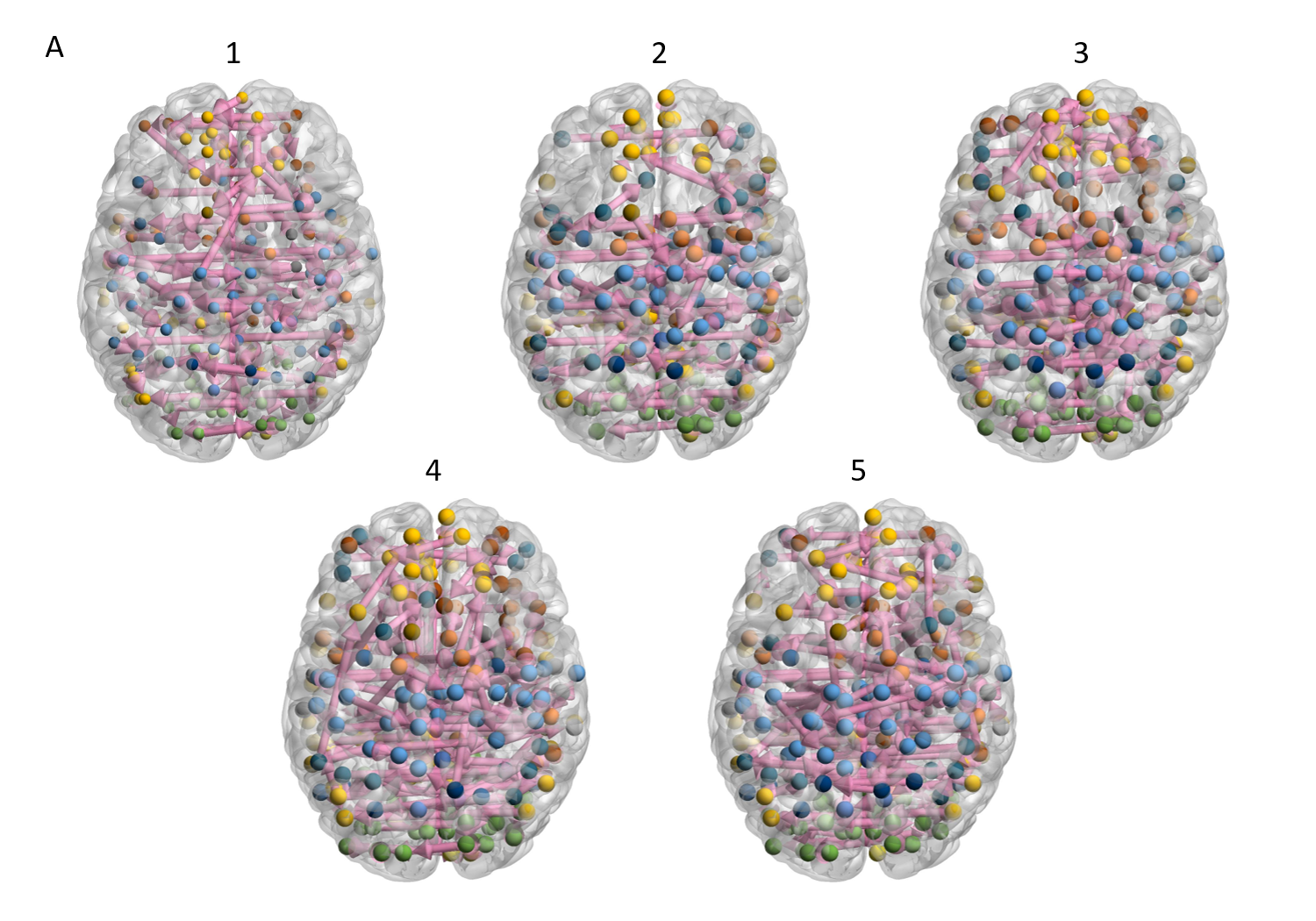}
\includegraphics[width=7.2in]{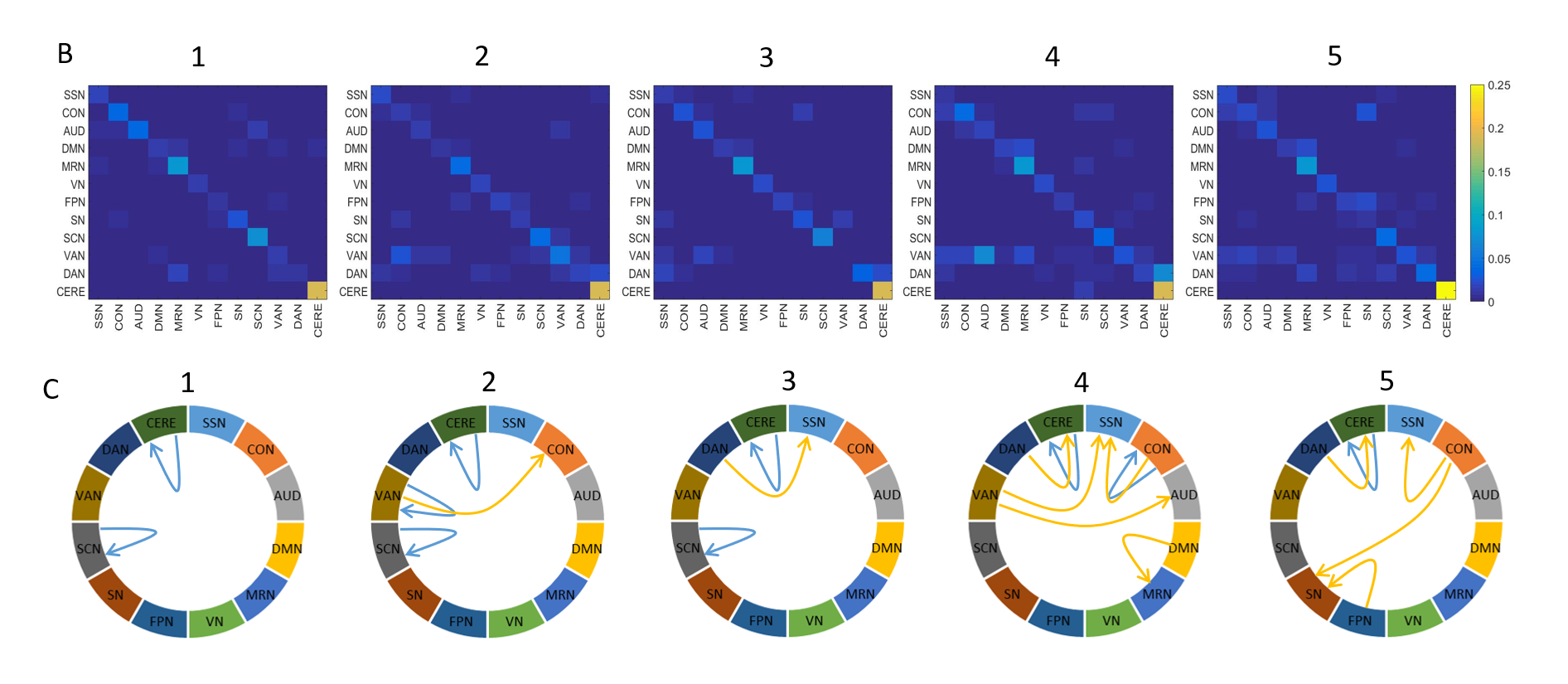}
\includegraphics[width=7.2in]{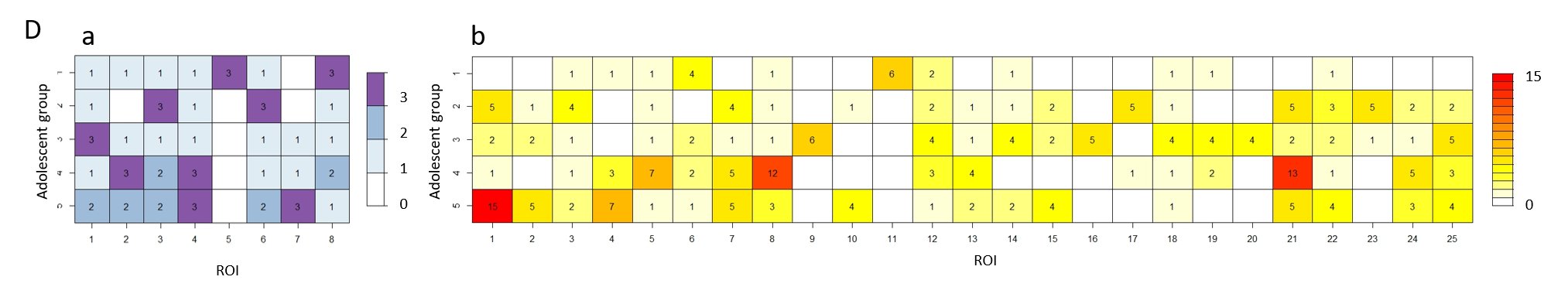}
\caption{Causal brain connectivity development from pre-adolescence to post-adolescence. The number index ($1$ to $5$) corresponds to the age category in Table \uppercase\expandafter{\romannumeral2}. A, Axial views of emotion-related node-level causal networks, where the arrows indicate the causal flow.  B, Heatmaps of the mean edge degrees, module-wise. C, Identified intra- (blue arrows) and inter- (yellow arrows) module causal flows. D, Development of emotion-related hubs (D.a: in-hubs, D.b: out-hubs) over various adolescent groups with detailed ROI information in Table \uppercase\expandafter{\romannumeral3} (for D.a) and \uppercase\expandafter{\romannumeral4} (for D.b).}\label{fig:3}
\end{figure*}


We examined intramodular and intermodular connectivity over the 5 adolescent groups. Fig \ref{fig:3}B visualized the average directed edge degrees within and across modules, where the rows indicate the beginning of the arrows and the columns indicate the end of the arrows. From Fig \ref{fig:3}B, the intra-module connectivity of DMN, SCN and CERE are strongly activated for all $5$ groups. As age increases, there is an increasing intermodular connectivity. We then conducted hypergeometric tests based on the number of edges module wise, and significant intra- and inter- causal connections are shown in Fig \ref{fig:3}C at significance level $\alpha = 0.05$ with FDR correction. The intramodular connectivity of the CERE were significantly activated of all adolescent groups for the emotion identification task, while the role of the SCN was only significant until the middle adolescent period. In addition, we found substantial intra-connectivity of SCN in the early adolescent group and CON in the late adolescent group. From the aspect of intermodular connectivity, no significant causal flows were found in the pre-adolescent group, $1$ each was identified for the early (VAN $\rightarrow$ CON) and middle (DAN $\rightarrow$ SSN) adolescent groups, $5$ were identified for the late adolescent group (CON $\rightarrow$ SSN, VAN $\rightarrow$ SSN, VAN $\rightarrow$ AUD, DMN $\rightarrow$ MRN, DAN $\rightarrow$ CERE), and $4$ causal flows were discovered in the post-adolescent group (CON $\rightarrow$ SSN, DMN $\rightarrow$ MRN, CON $\rightarrow$ SN, FPN $\rightarrow$ SN).

\subsubsection{Development of emotion-related hubs}

\begin{table}[h]
  \centering
  \setlength\tabcolsep{1.2pt}
  \caption{Anatomical location, functional network module and MNI coordinates of the identified in-hub ROIs.}
  \begin{threeparttable}
    \begin{tabular}{c|c|c|c|c}
    \hline
    ROI   & MNI (X, Y, Z) & Module & AAL   & Abbrev. \\
    \hline
    1     & 20,     -29,    60    & SSN   & Precentral Gyrus (R) & PG.R \\
    2     & 8,      -48,    31    & DMN   & Mid-cingulate gyrus (R) & MCG.R \\
    3     & -2,     -35,    31    & MRN   & Posterior cingulate gyurs (L) & PCG.L \\
    4     & 4,      -48,    51    & MRN   & Precuneus (R)  & PQ.R \\
    5     & -10,    11,     67    & VAN   & Supplementary motor area (L) & SMA.L \\
    6     & 36,     22,     3     & SN    & Insula (R)  & INS.R \\
    7     & 10,     22,     27    & SN    & Anterior cingulate gyrus (R) & ACG.R \\
    8     & 12,     -17,    8     & SCN   & Thalamus (R)  & THA.R \\
    \hline
    \end{tabular}%
 \begin{tablenotes}
    \footnotesize
    \item[*]  The ROI index corresponds to the row order from Fig. \ref{fig:3} D.a.
 \end{tablenotes}
\end{threeparttable}
\end{table}%

\begin{table}[h]
  \centering
  \setlength\tabcolsep{1.2pt}
  \caption{Anatomical location, functional network module and MNI coordinates of the identified out-hub ROIs. }
  \begin{threeparttable}
    \begin{tabular}{c|c|c|c|c}
    \hline
    ROI   & MNI (X, Y, Z) & Module & AAL   & Abbrev. \\
    \hline
    1     & -14    -18    40    & SSN   & Amygdala (L) & AMY.L \\
    2     & 29     -17    71    & SSN   & Precentral gyrus (R) & PG.R \\
    3     & -40    -19    54    & SSN   & Precentral gyrus (L) & PG.L \\
    4     & 19     -8     64    & CON   & Superior frontal gyrus (R) & SFG.R \\
    5     & -10    -2     42    & CON   & Mid-cingulate gyrus (L) & MCG.L \\
    6     & -13    -40    1     & DMN   & Precuneus (L)  & PQ.L \\
    7     & 15     -63    26    & DMN   & Precuneus (R)  & PQ.R \\
    8     & -2     -37    44    & DMN   & Mid-cingulate gyrus (L) & MCG.L \\
    9     & -10    55     39    & DMN   & Superior frontal gyrus (L) & SFG.L \\
    10    & -20    45     39    & DMN   & Superior frontal gyrus (L) & SFG.L \\
    \multirow{2}{*}{11}    & \multirow{2}{*}{13     30     59 }   & \multirow{2}{*}{DMN}   & Superior frontal gyrus,  & \multirow{2}{*}{SFGM.R} \\
    &&&medial (R)&\\
    12    & -26    -40    -8    & DMN   & Parahippocampus (L) & PHIP.L \\
    13    & 18     -47    -10   & VN    & Lingual gyrus (R) & LG.R \\
    14    & -15    -72    -8    & VN    & Lingual gyrus (L) & LG.L \\
    15    & 6      -72    24    & VN    & Cuneus (R) & Q.R \\
    16    & 11     -39    50    & SN    & Mid-cingulate gyrus (R) & MCG.R \\
    \multirow{2}{*}{17}    & \multirow{2}{*}{48     22     10}    & \multirow{2}{*}{SN}    & Inferior frontal gyrus,  & \multirow{2}{*}{IFGT.R} \\
    &&&triangular (R)&\\
    \multirow{2}{*}{18}    & \multirow{2}{*}{37     32     -2}    & \multirow{2}{*}{SN}    & Inferior frontal gyrus,  & \multirow{2}{*}{IFGO.R} \\
    &&&orbital (R)&\\
    19    & 26     50     27    & SN    & Middle frontal gyrus (R) & MFG.R \\
    20    & 9      -4     6     & SCN   & Ventral Anterior Nucleus (R) & VA.R \\
    21    & 52     -33    8     & VAN   & Superior temporal gyrus (R) & STG.R \\
    22    & 51     -29    -4    & VAN   & Middle temporal gyrus (R) & MTG.R \\
    23    & -52    -63    5     & DAN   & Middle temporal gyrus (L) & MTG.L \\
    24    & 46     -59    4     & DAN   & Middle temporal gyrus (R) & MTG.R \\
    25    & 29     -5     54    & DAN   & Precentral gyrus (R) & PG.R \\
    \hline
    \end{tabular}%
   \begin{tablenotes}
    \footnotesize
    \item[*]  The ROI index corresponds to the row order from Fig. \ref{fig:3} D.b.
 \end{tablenotes}
\end{threeparttable}
\end{table}%

To gain more insights into the affective emotion circuits and their development with age, we analyzed hub nodes for each group. Here we define hubs as the nodes with degrees at least two standard deviation higher than the mean degrees \cite{Jian2018}. We identified two types of hubs: in-hubs and out-hubs, which are selected through the in- and out-degrees, respectively. The in-degree of a node $i$ is defined as the number of directed edges that end at node $i$ (i.e. $\rightarrow i$), and the out-degree of a node $i$ is the number of directed edges that begin from node $i$ (i.e. $i \rightarrow$). The in-hubs can be treated as centers that receive information, while out-hubs are central nodes that convey out information. We identified hubs for each adolescent group, separately, and tracked their changes across groups.

Fig. \ref{fig:3}D.a and Table \uppercase\expandafter{\romannumeral3} give the in-hub development and the detailed in-hub information. The ROI at SMA.L has a pre-adolescence specific pattern. The ROIs at PQ.R and ACG.R have increased activities of receiving messages, especially, the ROI at ACG.R only starts to develop from middle adolescence. The remaining in-hubs have fluctuating trajectories. From Fig. \ref{fig:3}D.b and Table \uppercase\expandafter{\romannumeral4}, several out-hubs start to develop in a fluctuating manner after pre-adolescence, whose anatomical locations are at AMY.L, SFG.L (ROI 9, 10), PG.R, PQ.R, LG.R, Q.R, MCG.R, IFGT.R, VA.R, STG.R and MTG (ROI 23, 24). Some group-specific patterns have also been detected: the ROIs at SFGM.R for pre-adolescent group; the ROIs at IFGT.R and MTG.L for early adolescent group; the ROIs at SFG.L, MCG.R, VA.R for middle adolescent group; the ROIs at MCG.L (ROI: 5, 8), STG.R for late adolescent group; the ROIs at AMY.L and SFG.R for post-adolescent group.

\subsubsection{Discussion}

Our analysis revealed the developmental trajectory of directed brain circuitry during emotion identification tasks over various adolescent groups. We found intra- and inter- modular development, and pinpointed emotion-related hubs as well as several group-specific patterns. Our findings provide a causation template of emotion processing in the developing brain.

\emph{Intra-modular development}: We found a developmentally stable intra-modular activation anchored in the default mode (DMN), subcortical (SCN) and cerebellum (CERE) networks. The default mode network is important for mentalizing and inferring emotional states of others (\cite{blakemore2008social}); subcortical regions have a pivotal role in cognitive, affective, and social functions in humans \cite{koshiyama2018role}; the cerebellum contributes prominently to processing emotional facial expression \cite{Ferrucci2011}. The intra-modular activities of CERE increased significantly in the post-adolescent group, which may emerge in late puberty \cite{tiemeier2010}. The significance of intra-modular activities of the SCN appeared from pre-adolescence to middle adolescence, which has proven to be an important developmental period for subcortical brain maturation \cite{dennison2013}.

\emph{Inter-modular development}: As age increases, more intermodular connectivity emerges during emotion-related processing. Starting from early adolescence, inter-connections start to build among VAN, CON, SN and SSN. Specifically, SSN exhibits significant inter-connections of receiving information from other funtional network modules after middle adolescence. VAN plays an important role of conveying information in late adolescent group, and SN is crucial for receiving information in post-adolescent group. Two stable causal influences from CON to SSN, DMN to MRN become established after the late adolescent period. CON facilitates the maintenance of task-relevant goals and the incorporation of error information to adjust behaviors \cite{cocchi2013dynamic} and SSN (includes somatosensory cotex, motor regions and extends to the supplementary motor areas) is involved in performing and coordinating motor-related tasks like finger tapping. Gehringer et al. \cite{gehringer2019strength} proposed that maturation of the somatosensory system during adolescence contributes to the improved motor control. They further discovered that altered attenuation of the somatosensory cortical oscillations might be central to the under-developed somatosensory processing and motor performance characteristics in adolescents. Our results agreed with their conclusion and may provide a possible explanation. Besides, Sestieri et al. \cite{sestieri2011episodic} confirmed that responses in DMN regions peaked sooner than non-DMN regions during memory retrieval, and the parietal regions of DMN directly supported memory retrieval.

\emph{Emotion-related hubs}: As shown in Table \uppercase\expandafter{\romannumeral3} \& \uppercase\expandafter{\romannumeral4}, most identified hubs were located in the right hemisphere since it is dominant in the perception of facial expression and important for processing primary emotions \cite{alfano2008alteration}. Particularly, the hubs at PG, MCG, PQ, MTG play central roles in socioemotional processing. The precentral gyrus (PG) of the somatosensory cortex is related to recognition of facial and vocal expressions of emotion and a main effect of emotional valence on brain activity has been found in the PG.R \cite{seo2014neural}. A previous study \cite{shackman2011integration} verified that the mid-cingulate gyrus (MCG) is a hub linking incoming affective information with brain regions involved in goal-directed behavior. We further discovered that it is also a hub for distributing affective information. Precuneus (PQ) activation has been implicated in emotional and memory-loaded processes. The current study suggests that the PQ may play a direct role in the regulation of amygdala reactivity to emotional stimuli \cite{ferri2016emotion}, which explains its prominence as a out-hub location. Studies of emotional face recognition \cite{haxby2002human, sabatinelli2011emotional} identified the middle temporal gyrus (MTG) as a primary neural substrate for suprathreshold processing of the emotional expression of faces, which is consistent with our result of MTG as a central node to pass out information.

\emph{Developmental hub structures and group-specific hub patterns}: The majority of networks during development fluctuate, except for the steady increase of the in-hub activities at the PQ.R and ACG.R. Group-specific patterns have also been identified: the in-hub at SMA.L and out-hub at SFGM.R for pre-adolescence; the out-hubs at IFGT.R and MTG.L for early adolescence; the out-hubs at SFG.L, MCG.R, VA.R for middle adolescence; the out-hubs at MCG.L, STG.R for late adolescence; the out-hubs at AMY.L and SFG.R for post-adolescence. Some of our results have been previously supported in the literature. The role of developmental centers at PG, PQ, ACG, LING and PHIG remains consistent with our previous study of brain connectivity development in adolescence \cite{zhangTMI2019}. In this study, we further pinpointed their specific functions in the emotion circuit through directed graphical models. Another study of brain development from adolescence to adulthood \cite{kundu2018integration} also brought attention to age-related changes in the PQ. In \cite{simmonds2017protracted}, fluctuating trajectories in the MCG during adolescence were discovered.

\begin{figure}[h!]
\centering
\includegraphics[width=3.5in]{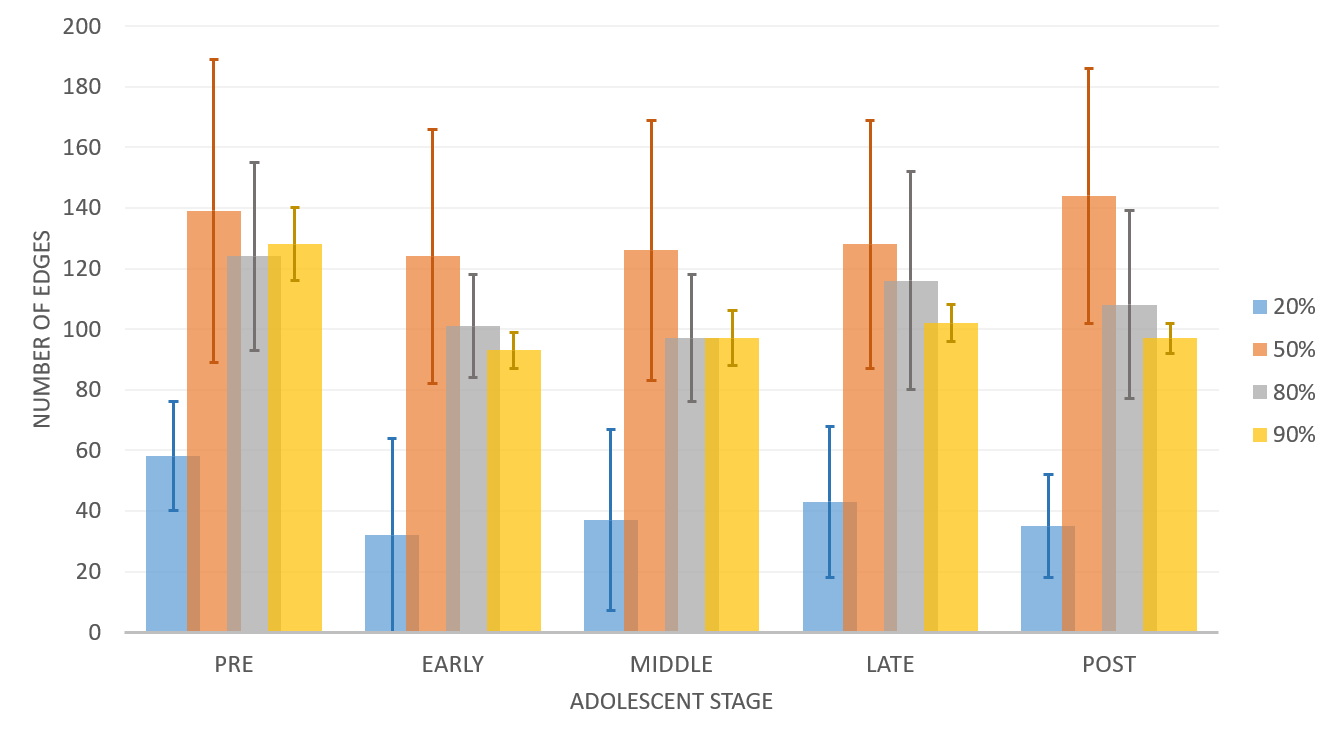}
\caption{Robustness of BiLiNGAM with various percentages ($20 \%, 50 \%, 80 \%, 90 \% $) of samples ($N = 855$) from the PNC dataset. Data were mean number of edges $\pm$ stardard deviation.}\label{fig:4}
\end{figure}

\emph{The effect of sample size and robustness}: We examined the robustness of our proposed method and assessed the effect of sample size on robustness for each group. We randomly drew $m$ percent of $N$ participants ($m \in \{20 \%, 50 \%, 80 \%, 90 \% \}$, $N = 855$) from the PNC dataset, sample without replacement and with the proportion of the $5$ adolescent groups. We then applied the BiLiNGAM method with the subsamples. For each sample size, we repeated the procedure $10$ times. In Fig. \ref{fig:4}, we showed the mean number of edges detected for each adolescent group under various sample sizes. From pre-adolescence to post-adolescence, the trajectory of each sample remains similar. However, we found that a limited number of edges were identified with small sample size ($m=20\%$). As the sample size increases, the identified number of edges for each group becomes steady and the variance decreases significantly.

\section{Conclusion}

In this paper, we propose a multiple DAG estimation for non-Gaussian data, with the name of Bayesian incorporated linear non-Gaussian acyclic model (BiLiNGAM) in the sense that we were enlightened by the idea of Bayesian prior and implemented a joint Bayesian-incorporating estimation to acquire the priors. The main contributions of our work can be summarized as follows. First, from a mathematical perspective, BiLiNGAM was, to the best of our knowledge, the first method to jointly estimate related DAG models in the high-dimensional setting for non-Gaussian data. Second, our proposed method accomplished the integration of undirected graph with directed acyclic graph, in the sense that we incorporated the undirected graph estimation as prior information into the direct LiNGAM model to perform DAG construction. In other words, we use the undirected graphs to mitigate the irrelevant information to facilitate casual inferences, which speeds up numerical convergence and computation. Third, the simulation results in Section \uppercase\expandafter{\romannumeral3} show that BiLiNGAM can maintain a stable and accurate performance over various settings. In particular, the proposed BiLiNGAM is superior for high dimensional cases. Finally, the analysis of brain's emotion circuit development revealed the trajectory of directed brain circuitry during emotion identification tasks over various adolescent groups. We have found several significant intra- and inter- modular networks that change over developmental stages, and pinpointed emotion-related hubs as well as various group-specific patterns. Our findings provide a causation template of emotion processing in the developing brain.

\section*{Acknowledgment}
The work has been funded by NIH (R01GM109068, R01MH104680, R01MH107354, P20GM103472, 2R01EB005846, 1R01EB006841, R01MH121101, R01MH116782, R01MH118013, P20GM130447), and NSF (\#1539067).

\bibliographystyle{IEEEtran}
\bibliography{refer}

\end{document}